\documentclass{llncs}

\usepackage{microtype}
\usepackage{graphicx}
\usepackage{subfigure}
\usepackage{booktabs} 

\usepackage{hyperref}

\usepackage{pslatex}
\usepackage{soul}
\usepackage{csvsimple}

\usepackage[textsize=tiny]{todonotes}

\usepackage{listings}
\usepackage{xcolor}
\definecolor{codegreen}{rgb}{0,0.6,0}
\definecolor{codegray}{rgb}{0.5,0.5,0.5}
\definecolor{codepurple}{rgb}{0.58,0,0.82}
\definecolor{backcolour}{rgb}{0.95,0.95,0.92}

\lstdefinestyle{cstyle}{
    language=C,
    basicstyle=\ttfamily\small,
    backgroundcolor=\color{backcolour},   
    commentstyle=\color{codegreen},
    keywordstyle=\color{magenta},
    numberstyle=\tiny\color{codegray},
    stringstyle=\color{codepurple},
    breaklines=true,
    captionpos=b,                    
    keepspaces=true,                 
    showspaces=false,                
    showstringspaces=false,
    showtabs=false,                  
    tabsize=2,
    moredelim=[is][\color{red}]{@@}{@@},
    }

\lstdefinestyle{csvstyle}{
    basicstyle=\ttfamily\footnotesize,
    columns=fullflexible,
    breaklines=true,
    postbreak=\mbox{\textcolor{red}{$\hookrightarrow$}\space},
    frame=single,
    captionpos=b,
    showstringspaces=false,
    moredelim=[is][\textbf]{**}{**}
}

\lstdefinestyle{cotstyle}{
    basicstyle=\normalsize\ttfamily,
    language={},
    moredelim=[is][\color{red}]{@@}{@@}
}

\usepackage{mdframed}
\newmdenv[
    linecolor=black,
    linewidth=2pt,
    backgroundcolor=gray!20,
    frametitlebackgroundcolor=gray!20,
    frametitlerule=true,
    roundcorner=10pt
]{mybox}

\definecolor{babyblue}{rgb}{0.54, 0.81, 0.94}

\begin{document}

\pagestyle{plain}

\title{%
Specify What? 
Enhancing Neural Specification Synthesis by Symbolic Methods\thanks{This work was supported by the Wallenberg AI, Autonomous Systems and Software Program (WASP), funded by the Knut and Alice Wallenberg Foundation.}}

\titlerunning{Specify What? Enhancing Neural Specification Synthesis by Symbolic Methods}

\author{George Granberry \orcidID{0009-0005-4628-5464} \and 
        Wolfgang Ahrendt\orcidID{0000-0002-5671-2555} \and 
        Moa Johansson\orcidID{0000-0002-1097-8278}
        }
\institute{Chalmers University of Technology, Gothenburg, Sweden \\
\email{\{georgegr, ahrendt, jomoa\}@chalmers.se}
}
%

%

%
\maketitle


\begin{abstract}
We investigate how combinations of Large Language Models (LLMs) and symbolic analyses can be used to synthesise specifications of C programs. The LLM prompts are augmented with outputs from two formal methods tools in the Frama-C ecosystem, Pathcrawler and EVA, to produce C program annotations in the specification language ACSL. We demonstrate how the addition of symbolic analysis to the workflow impacts the quality of annotations: information about input/output examples from Pathcrawler produce more context-aware annotations, while the inclusion of EVA reports yields annotations more attuned to runtime errors. In addition, we show that the method infers the programs intent, rather than its behaviour, by generating specifications for buggy programs and observing robustness of the result against bugs.


\end{abstract}

\section{Introduction}
The field of specification synthesis offers a possible solution to the inherent complexities involved in creating and maintaining specifications for software verification. Creating useful specifications demands a deep understanding of both the specification language and the verification process, which can often be as intricate, if not more so, than the software they aim to verify. This complexity poses a significant barrier \cite{davis2013study,tyler2021formal}, especially in dynamic environments where frequent updates and refactoring are the norm. Maintaining an accurate alignment between ever-evolving code and its specifications can become a cumbersome and error-prone process.

Specification synthesis potentially alleviates these concerns by automating the generation and adaptation of specifications. Instead of requiring developers to manually write detailed specifications -- a task that can be both time-consuming and susceptible to human error -- specification synthesis aims to infer and edit specifications directly from the codebase and associated context. The goal is to transform specifications into convenient guardrails that provide valuable insights and guidance to programmers, rather than chores performed at the end of the software pipeline.

Approaches to generating specifications typically employ a range of symbolic techniques, encompassing static as well as dynamic analyses \cite{lathouwers2024survey}. For instance, Daikon \cite{ernst2007daikon}, a widely recognised tool in dynamic analysis, infers properties by observing program behaviour at runtime. On the other hand, static analysers deduce properties based on the program's structure without executing it. Despite their precision, the primary limitation of these methods is their rigidity. Symbolic techniques are constrained by a limited range of expressible properties and typically specialise in specific types of analyses which restricts their flexibility in adapting to diverse verification needs.

On the other side of specification synthesis techniques are machine-learning-based Natural Language Processing (NLP)~\cite{blasi2018translating} and Large Language Models (LLMs) \cite{brown2020language}. NLP tools specialise in understanding human language, such as comments, while LLMs stand out for their flexibility and creativity in when dealing with arbitrary inputs. These models can theoretically generate any specification that can be articulated in their associated language, provided that they are appropriately trained and given the correct prompts.

However, this strength also introduces a significant challenge: the large range of potential specifications LLMs can produce often includes outputs that may not be practically useful or even plain wrong. While an LLM can generate a wide array of specifications, the lack of inherent direction means that there is no guarantee that the generated specifications will be relevant or valuable for specific verification tasks. This challenge has led users of LLM-based synthesis to rely on \emph{prompt engineering} \cite{white2023prompt} techniques in order to increase the likelihood of the LLM to produce specifications that align with their objectives.

In this paper, we introduce a hybrid approach that combines the precision of existing symbolic tools with the flexibility and creative potential of LLMs. By integrating outputs from symbolic analysis of C programs into LLM prompts, this method aims to harness the generative capabilities of LLMs while taking into account the focus and direction of symbolic analysis. As interpreting specifications is subjective, we rely on a human-in-the-loop qualitative analysis to observe patterns in our generated specifications. From a practical sense, we cannot always expect our code to be semantically correct when generating a specification for it. Therefore, we also investigate how our proposed technique interacts with incorrect code. Ultimately, we want specification generation to contribute to the revealing of bugs in the code.

We observed that adding the output of symbolic analyses to the input of the LLM reduces the number of generated specifications. Instead, it increases the quality and the level of abstraction, such that the specification represents more intention and less low-level details.
The integration acts as a directive lens, focusing the LLM on generating specifications that align with insights from symbolic analysis, thus yielding specifications that are more relevant to the user who chose said symbolic tools. Each symbolic analysis is interpreted and utilised in a unique way.
In general, the analysis tools provide some extra context, and increase the likelihood that the LLM will focus on particular aspects of the behaviour.

The dominating factor for the quality of generated specifications seems to be the extent to which a program's \emph{intent} can be identified.  In the absence of a clearly inferred intent, the LLM tends to default to generating specifications based on low-level implementation details or the provided context. But whenever the LLM (partly with help of symbolic pre-analysis) successfully grasps the purpose of the program that a programmer \textit{intended} to write, it becomes much more likely to produce specifications that align closely with that intent. This is not only true for programs which match their intent, but even for buggy programs. In our experiments, the LLM was more attentive to the intended behaviour than to the actual behaviour, as generated specifications were rather robust against bugs in the programs. This is very encouraging, and emphasizes the relevance of our approach and further work in this direction. The more we can automatically generate specifications which reflect the intended behaviour of a program, the more this can help us to reveal bugs, i.e., identify deviations from the intended behaviour.

\section{Tools and Languages}

\subsubsection{Frama-C and ACSL}
The Frama-C ecosystem is an open-source suite of tools designed for the analysis of the source code of software written in C \cite{Farma-C,frama-cBook2024}. It integrates various static and dynamic analysis techniques to evaluate the correctness, safety, and security of C programs. It also supports the specification language ACSL \cite{baudin2008acsl,e_acsl_reference}, which is used to formulate \emph{contracts} consisting of, among others, preconditions – assumptions on the input and prestate of a function – and postconditions – requirements on the output and poststate of a function.
These contracts, examples of which can be seen both in Sec.\ref{section:qual} or the appendix, provide a clear and formal framework for understanding and
verifying a function’s behaviour. Other ACSL annotations commonly used are \emph{assertions} - stating a condition that needs to be true at some point in execution - and \emph{loop invariants} which specify conditions that need to be maintained by every iteration of a loop.

\subsubsection{Automated Testing: Pathcrawler}
The PathCrawler tool is designed for the automated testing of C programs \cite{williams2005pathcrawler}. Its primary function is to generate and execute test inputs for C code, with a particular focus on achieving high code coverage. Employing a technique known as concolic testing \cite{concolicTestingC05} -- a combination of concrete and symbolic testing -- Pathcrawler efficiently explores different execution paths in the program. First it generates test inputs and then executes them, providing valuable information from the execution results across a broad spectrum of program paths. In addition, PathCrawler allows users to incorporate a test oracles, classifying the outcome of every test case of some function. However, we want to highlight that we did not make such oracle implementations available to the LLM when asking it to generate specifications. Generic oracles can be seen as executable specifications, and would have diluted the significance of our experiments. Instead, we only include input/output pairs, with the non-generic verdict.

\subsubsection{Value Analysis: EVA}
The EVA static analyzer uses abstract interpretation to approximate a set of possible values that program variables can take to avoid certain runtime errors \cite{EVA}. By doing so, it can
identify a range of potential issues, such as division by zero, buffer overflows, null
pointer dereferences, and arithmetic overflows. EVA’s analysis helps in ensuring
that the code behaves correctly across all possible execution paths and input
values. EVA is designed to respect, and work with, ACSL annotations when they are present.

\subsubsection{LLM and Prompts}
We have chosen to use GPT-4 \label{gpt4} (version gpt-4-0125-preview) as our LLM for generating specifications. We ran preliminary tests with Gemini as well as GPT-3.5 but found that they returned too many syntactical and semantic errors to draw interesting conclusions from. While open source models such as Llama-3 have recently gained traction, the setup and fine tuning of such a model for our purposes remain as future work.

We prompt GPT-4 with a C program, providing instructions for how to generate ACSL annotations in a step-by-step manner. We also include a few examples of valid annotations in the prompts, leveraging a form of “few-shot learning”~\cite{brown2020language} to guide the model (see Appendix \ref{PromptBaseline})\footnote{Note for reviewers: We include appendices for convenience. In a final version, references to appendices would be replaced by references to online material.}. Additionally, we employ “Chain-of-Thought”~\cite{wang2023chain} reasoning, prompting the LLM to explain its thought process (seen on pages\pageref{listing:cot1} and \pageref{listing:cot2}) for generating the annotations in the same output as the actual specification. Finally, we add an "Emotional Stimulation"~\cite{li2023large} instruction at the end of each prompt. For the core of the experiment, we augment the baseline prompt with outputs from the Pathcrawler and EVA tools to further aid the model (see Appendix \ref{PromptPC} and \ref{PromptEVA}).

\section{C-program Test Suites}
For our study, we have chosen to utilise the 55 programs from the Pathcrawler test suite, which we will refer to as \textbf{pathcrawler\_tests}. Note that the test suite is not available online; it was provided to us by the Pathcrawler developers. Thereby, we can assume that this test suite was not directly used in the LLM's training (although it might have seen similar ones). This suite includes a variety of program types, balancing well-known algorithms like Binary Search with more niche programs such as a Soup Heater controller. It also contains small, specially crafted programs designed to test specific capabilities of Pathcrawler, adding another layer of diversity to our tests.
According to the Pathcrawler developers, the 55 programs are supposedly correct, in the sense that they are believed to correspond to their respective intention, and have no known bugs. Consequently, we examine with this suite to which extent our method produces accurate annotations for supposedly correct programs.

To also investigate to which extent our approach can help with buggy programs, we created a second suite of programs titled \textbf{mutated\_set}. This comprises 8 programs from \textbf{pathcrawler\_tests} but with handcrafted mutations simulating typos, designed to explore a range of programs across two key dimensions: clarity of intent and complexity. In order to study the interactions between these dimensions, this set includes various types of programs: simple programs with clear intent, complex programs with clear intent, simple programs with less clear intent, and complex programs with less clear intent.

\section{Generating Annotations}
 For each program in our two test suites, we generate three sets of ACSL annotations: 
 \begin{enumerate}
     \item \textbf{baselines\_set}: Specifications generated using just the program in the prompt (Appendix \ref{PromptBaseline})
      \item \textbf{pathcrawler\_set}: Specifications generated by including a compact representation of test-cases generated by Patchcrawler in the prompt shown in Appendix \ref{PromptPC}.
     \item \textbf{eva\_set}: Specifications generated by running EVA on the program and including its report on potential value errors in the prompt shown in Appendix \ref{PromptEVA}.
     
 \end{enumerate} 

The variability of LLMs like GPT-4 can be adjusted via its "temperature" setting which controls the level of determinism during generation. As we are interested in exploring what the \textit{average} specification generated by a given prompt is, we choose to generate three distinct specifications for each program (and prompt) within our test suite, repeating the steps above with a temperature setting of 0.7. This approach allows us to both capture a spectrum of possible specifications as well as assess the consistency and variability of the model's output across multiple generations, while not being too economically costly.

Loop invariants are key for functioning specifications, and our specification generation process does indeed generate them. However, generating loop invariants is not the focus of this paper. The generation of loop invariants with LLMs calls for sophisticated prompts dedicated to this task \cite{kamath2023finding,flanagan2001annotation}. Rather, we focus on function contracts -- particularly preconditions, postconditions, and assigns clauses -- and our prompts are designed to generate these.
 
\section{Evaluation}
Evaluating specifications is challenging due to the absence of a definitive specification for any given program. Different users often have varying priorities and perspectives on which properties are worth verifying, making the notion of a definitive specification subjective. Similarly, a specification might be logically correct, but more or less trivial with respect to the program at hand, in which case it provides little value.

In light of these challenges, our evaluation methodology does not attempt to benchmark the generated specifications against a predefined gold standard, nor does it aim to determine the optimal approach to creating specifications. Instead, our focus is on identifying the behaviours and patterns that emerge from incorporating symbolic analysis outputs into the specification generation prompts. This approach allows to better understand the dynamics at play and what kinds of output to expect given particular prompt characteristics. In particular, we want to understand the impact of symbolic-analysis based prompt composition on the generated specifications.

We perform quantitative as well as qualitative evaluation from two different angles:
\begin{itemize}
    \item \textbf{Types of Annotations}: First, we count the number of each type of annotation that we get for each specification set.
    \item \textbf{Qualitative Analysis}: Additionally we use a human-in-the-loop qualitative analysis to interpret the specification and identify trends depending on which prompt composition technique was used, to assess how the different symbolic tool outputs influence the results of the LLM. For this we use the programs in the \textbf{pathcrawler\_set}.
    \item \textbf{Implementation vs. Intent}:  Finally, we specifically examine programs in the \textbf{mutated\_set} to study how errors introduced into the program affect the resultant specifications. This analysis explores how errors, symbolic analyses, intended program functionality, and actual implementation interact in the generation of specifications.
\end{itemize}

\subsection{Types of Annotations}
In this section we quantify the annotations generated across all programs for our three sets: \textbf{baseline\_set}, \textbf{pathcrawler\_set}, and \textbf{eva\_set}. While this approach does not say anything about the quality or semantics of any annotations, it does provide us with a macro-level view of which kinds of annotations are being generated and at what frequency. This systematic approach allows us to capture the influence of different symbolic contexts on the annotation generation process. It is important to note, however, that a larger number of annotations does not necessarily signify a better specification.

\begin{figure*}[htbp]
\includegraphics[width=\textwidth]
{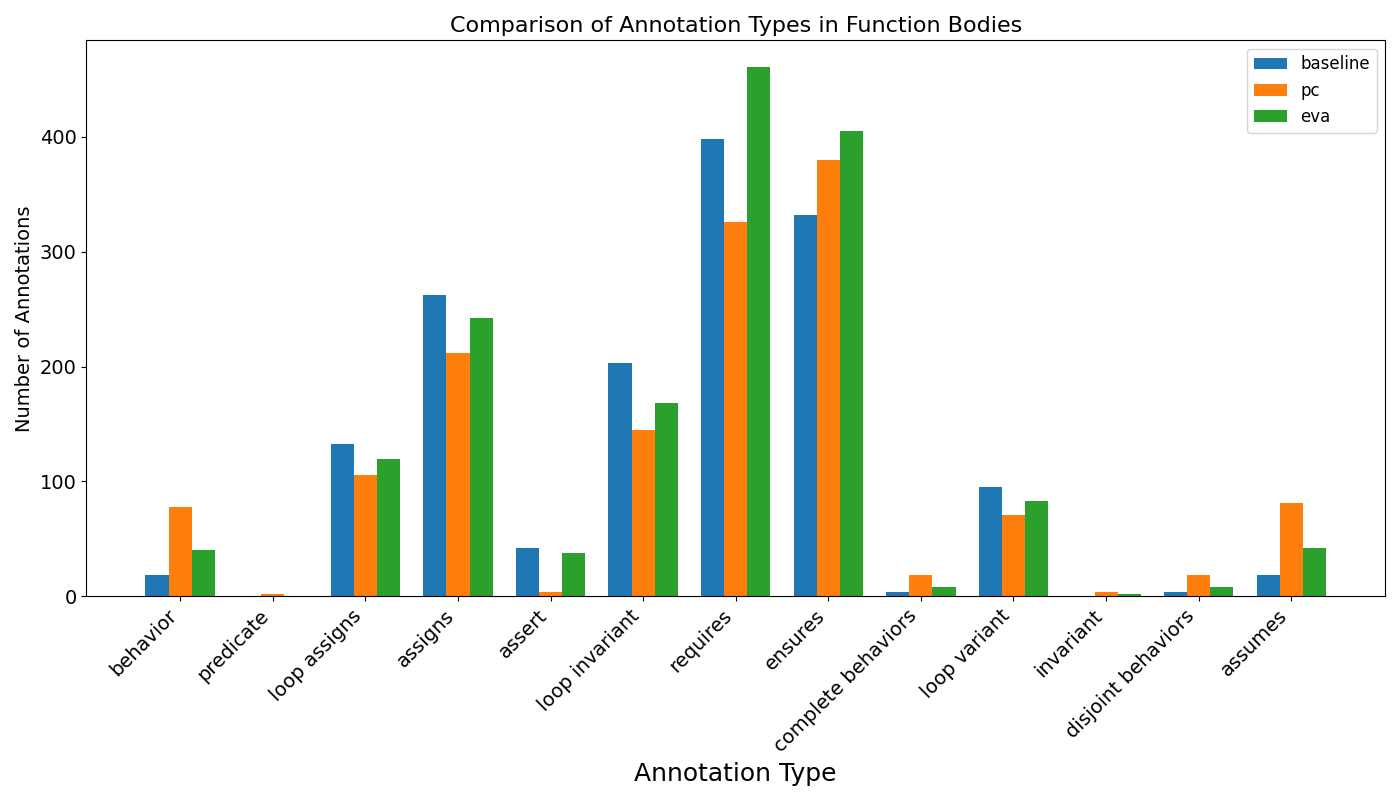}
\caption{Annotation-type counts for each prompt}
\label{fig:ifm_bar_counts.png}
\end{figure*}

Fig.\ref{fig:ifm_bar_counts.png} displays the number of annotations generated for each annotation type for the three prompts. For all three cases, the most common annotations are \emph{requires} and \emph{ensures} clauses, which are used to define pre- and post-conditions of functions, followed by \emph{assigns} statements and \emph{loop invariants}. More advanced ACSL annotations such as predicates or ghost-code are rarely, if ever, generated.

We observe that \textbf{eva\_set} has the largest number of preconditions and postconditions, hinting that EVA reports consistently provide information that can be translated into annotations. This lines up with the contents of EVA reports as they contain both alarms that define preconditions as well as a value analysis that describes all possible output values. The report and how it is utilised will be explored further in the upcoming qualitative analysis.

We also observe that \textbf{pathcrawler\_set} contains the largest number of behaviour clauses. In our experience with observing ACSL generated by GPT-4, the presence of behaviour clauses is generally an indication that the LLM has found non-trivial properties to add to its specification.

\subsection{Baseline prompt}
\label{section:qual}

Many of the annotations produced with the baseline prompt, which includes no additional symbolic context, were rather simplistic. While not necessarily incorrect or completely useless, these specifications tended to focus on surface-level details of the programs, overlooking more substantive aspects. This can be seen in Fig.~\ref{code:baseline_bsearch} which shows a specification generated for an implementation of binary search. The specification describes \textit{some} properties of the program, but completely misses the semantics of a search algorithm.

\begin{figure}[htbp]
\begin{lstlisting}[style=cstyle]
/*@
  @ requires \valid(A + (0..9)) && \valid(&elem);
  @ ensures \result == 0 || \result == 1;
  @ assigns \nothing;
*/
int testme( int A[10], int elem) {
  ...
  /*@
    @ loop invariant high >= low;
    @ loop assigns low, high, mid, ret;
    @ loop variant high - low;
  */
  while( ( high > low ) )
  ...
}
\end{lstlisting}
\caption{Specification generated for \textbf{binary search} program with the baseline prompt}
\label{code:baseline_bsearch}
\end{figure}

\subsubsection{CoT Inspection}
Our initial intuition was that the LLM was unable to infer that the code in Fig.\ref{code:baseline_bsearch} was an implementation of binary search, however by recording the CoT process we found evidence to the contrary. For example, the following snippet that was generated along with the annotated program:

\vspace{10pt}
\begin{mybox}
The program seems to implement a binary search algorithm on the array A[]. It tries to find a given element elem in the array A[] and returns a boolean indicating if it was found or not.
\end{mybox}
\vspace{10pt}

Despite this inference made by the LLM, the specification produced did not reflect its understanding. This pattern shows up frequently where the intent is shown to be \textbf{inferred} by the LLM but not given directly to it. For example the act of naming the function \textit{Bsearch} instead of \textit{testme} noticeably increases the quality of the specification.

Based on the relationship between what the LLM is able to infer in the CoT and the specifications it outputs, it suggests that a weak or limited specification does not necessarily reflect the inference or logical capabilities of the LLM. Rather, it indicates that the LLM was not provided with sufficient direction or emphasis on the aspects that we were looking for.

\subsection{Prompt with Pathcrawler Input/Output Pairs}
In conducting a qualitative analysis on \textbf{pathcrawler\_set}, we explore the efficacy of incorporating Pathcrawler input/output pairs into specification synthesis prompts. These input/output pairs are represented by a CSV string that includes the program input parameters, the output, and an optional “verdict” provided by an oracle. The rationale for this integration is grounded in the abstraction level that input/output pairs represent, which is among the most abstract representations of a program’s behaviour. An example of such a CSV string can be seen in Fig.\ref{csv:adpcm}, which represents the inputs and outputs for an Adaptive Differential Pulse Control Modulation (ADPCM) function. Given that these pairs are utilised for synthesising programs~\cite{gupta2020synthesize}, extending their use to specification synthesis is a logical progression.

\begin{figure}[hbpt]
\begin{lstlisting}[style=csvstyle]
input_n,input_valeur,input_t[0],input_t[1],output,verdict
2,0,-37,0,0,unknown
2,-91,0,62,0,unknown
2,0,0,12,1,unknown
\end{lstlisting}
\caption{CSV string representing inputs and outputs generated for Adaptive Differential Pulse Control Modulation (ADPCM) by Pathcrawler}
\label{csv:adpcm}
\end{figure}

\subsubsection{Abstract Thinking}
Our observations indicate that incorporating Pathcrawler input/output pairs into the specification generation process can significantly assist LLMs in achieving specifications with more abstract properties. By “abstract properties,” we mean characteristics that describe the purpose and high-level functionality of the program, rather than focusing on specific implementation details.

With the inclusion of Pathcrawler input/output pairs we observe that the LLM is able to understand the purpose of the program based on examples. Take for example the following snippet from the CoT process that we captured from the LLM during its specification generation process using Pathcrawler.

\begin{lstlisting}[style=cotstyle]
Next, let's analyze the PathCrawler CSV.

1. The CSV file contains test cases for `testme`. Each row represents a test case, with columns representing the inputs `len`, `input_buffer`, and `input_msg`, and the output.

2. @@From the test cases, we can observe that if `msg[i]` is `\n` and `msg[i+1]` is `.`, then `buffer[j]` gets `msg[i]`, `msg[i+1]`, and an extra `.`. Otherwise, `buffer[j]` simply gets `msg[i]`. This behavior could be represented as an ACSL annotation in a postcondition.@@
\end{lstlisting}
\label{listing:cot1}

Specifically highlighted in red, we see that the LLM noticing a pattern in the inputs and suggest a postcondition to represent it. This postcondition is then directly translated into an ACSL annotation for the specification shown in Fig.\ref{code:bugkpath_pc}. This pattern is repeated throughout the comparison between \textbf{baseline\_set} and \textbf{pathcrawler\_set} where the test pairs will bridge some sort of logical gap that makes the LLM more likely to specify abstract properties.

\begin{figure}[htbp]
\begin{lstlisting}[style=cstyle]
/*@ requires \valid_range(msg, 0, len-1);
  @ requires \valid_range(buffer, 0, BUFSZ-1);
  @ requires len >= 0 && len <= BUFSZ;
  @ assigns buffer[0..BUFSZ-1];
  @ ensures \forall integer i; 0 <= i < len ==>
  @            ((msg[i] == '\n' && msg[i+1] == '.') ?
  @              buffer[i] == msg[i] && buffer[i+1] == msg[i+1] && buffer[i+2] == '.' :
  @              buffer[i] == msg[i]);
*/
void testme (char *msg, int len, char *buffer)
{...}
\end{lstlisting}
\caption{BugKPath - Pathcrawler}
\label{code:bugkpath_pc}
\end{figure}

\subsubsection{Over-approximation}
Pathcrawler, as a structural test case generator, is designed to generate the minimal number of test cases necessary to cover all execution paths in a program. While a compact number of test cases can be considered a feature in some cases, this feature can also cause the LLM to produce over-approximated specifications. In this context, an over-approximated specification is one that makes assumptions about inputs and outputs based on a limited number of test-cases.

This behaviour can be seen in the specification generated for the program called \textbf{PointeurFonction5}, seen in Fig.\ref{code:pf5_pc}. In this example, Pathcrawler only providing two test cases leads to a specification overly tailored to the examples.

\begin{figure}[htbp]
\begin{lstlisting}[style=cstyle]
...
/*@
  @ requires a >= 0 && a <= 1;
  @ requires b >= 0 && b <= 1;
  @ assigns \nothing;
  @ ensures (\result == 2 && a == 0 && b == 0) || (\result == 1 && a == 1 && b == 0);
*/
unsigned int testme( int a, int b){...}
\end{lstlisting}
\caption{PointerFunction5 - Pathcrawler}
\label{code:pf5_pc}
\end{figure}

Notice how the preconditions specified limit our inputs to $\forall a,b \in \{0,1\}$. From the implementation of the program there doesn't seem to be any specific reason why the LLM should set that precondition. However, by looking into the CoT captured during specification generation we see a clear reasoning for why the precondition was included.

\begin{lstlisting}[style=cotstyle]
1. In the functions f0, f1, and f2, I add a postcondition (`ensures`) to specify the return value behavior since these functions return an integer value that is based on their input. This will allow better verification of the function behavior.
@@2. In the `testme` function, I add preconditions (`requires`) to limit the range of the input parameters `a` and `b` to prevent overflow or underflow issues. The range is derived from the PathCrawler output.@@
3. I add postconditions (`ensures`) to specify the expected return value behavior of the `testme` function based on the PathCrawler output.
4. I also add `assigns` clause to the `testme` function to specify which variables the function is allowed to modify.
\end{lstlisting}

The text highlighted in red clearly indicates that the preconditions were derived from the input/output pairs generated by Pathcrawler. Therefore, we should be able to look at the Pathcrawler input/output pairs to understand why the LLM came up with those preconditions. Upon examining the Pathcrawler output, we notice that there are only two test cases: one where $a=0$ and one where $a=1$. These inputs are not necessarily related to any of the semantics of the code, but the LLM overgeneralised based on the small number of examples.

\begin{lstlisting}[style=cotstyle]
input_a,input_b,output,verdict
0,0,2,unknown
1,0,1,unknown
\end{lstlisting}

\subsubsection{State Mutation}
The number of examples is not the only criterion for improved outcomes. The LLM also needs to detect recognisable patterns within these examples; without such patterns, even a large set of test cases may not lead to significant improvements in specification quality. Take for example the program titled \textbf{Apache} which edits URL strings. Based on its large number of test cases shown below, one would expect this large number of test cases to produce more useful annotations. However at a closer look, this program produces no return value and instead manipulates state.

\begin{lstlisting}[style=csvstyle]
input_scheme,input_uri[0],..,input_uri[14],output,verdict
1,47,47,0,0,0,0,0,0,0,0,0,0,0,0,0,,unknown
1,47,58,0,0,0,0,0,0,0,0,0,0,0,0,0,,unknown
2,108,47,47,47,0,0,0,0,0,0,0,0,0,0,0,,unknown
...
422214939,47,0,0,0,0,0,0,0,0,0,0,0,0,0,0,,unknown
5,108,100,97,112,47,47,47,63,63,63,0,0,0,0,0,,no_extra_coverage
\end{lstlisting}

Because there is no output to relate the input to, the LLM is unable to make any conclusions about the CSV file. Rather than it just not providing any benefit, we found that such additions to the prompt were actually detrimental to specification generation. This could possibly be because the CSV file takes attention away from the provided program, diverting the LLM’s focus and thereby reducing the overall quality.

\subsubsection{Are Test Cases Useful?}
As discussed in previous sections, adding Pathcrawler input/output pairs to prompts for specification generation is not a silver bullet that will always improve the quality of specifications. This observation raises the question: does adding input/output examples to prompts have value in general?

It is important to distinguish between the weaknesses of a specific tool and the limitations of the methodology itself. In this context, our two main weaknesses are: a small number of test cases and no representation of how state changes during test execution. These weaknesses are features of Pathcrawler (which was not designed with our use case in mind) rather than inherent flaws in the concept of using test cases for specification generation.

To address these weaknesses, we could develop a test case generator that overcomes these limitations. Such a generator would need to produce a larger quantity of test cases and include the result of state changes from before and after execution. Based on our observations, we believe that a suitable test case generator, free from these constraints, could be beneficial in increasing the quality of the generated specifications.


\subsection{Prompt with EVA annotations}
EVA, a static analysis tool, specialises in conducting a comprehensive value analysis of C programs. The report that it generates includes the detailed outcomes of the value analysis but also lists alarms that signify possible runtime errors linked to these value states. Such alarms are indicators of conditions under which the program might fail or behave unexpectedly, essentially flagging areas of the code that are prone to errors due to specific input values or execution paths.

\subsubsection{Avoiding Runtime Errors}
In our analysis of the \textbf{eva\_set}, a distinct characteristic emerges: the prevalence of precondition annotations that serve to help the function avoid runtime errors. These preconditions appear to be a direct result of the runtime alarms included the EVA reports. The alarms detail problematic input ranges, such as values that could cause the program to behave unpredictably or fail -- common issues highlighted include index-out-of-bound errors, divide-by-zero errors, and integer overflows.

An example of these value domains can be seen in the specification generated for the program \textbf{Alias5}, seen in Fig.\ref{code:alias5_eva}. The specification clearly aligns with the EVA report included in its synthesis prompt.

\begin{lstlisting}[style=cotstyle]
...
[eva:alarm] temp_files/tmphpdyn83w/eva_temp.c:8:Warning:
  signed overflow. assert -2147483648 <= x * 2;
[eva:alarm] temp_files/tmphpdyn83w/eva_temp.c:8:Warning:
  signed overflow. assert x * 2 <= 2147483647;
[eva:alarm] temp_files/tmphpdyn83w/eva_temp.c:9:Warning:
  signed overflow. assert -2147483648 <= v - y;
[eva:alarm] temp_files/tmphpdyn83w/eva_temp.c:9:Warning:
  signed overflow. assert v - y <= 2147483647;
[eva:alarm] temp_files/tmphpdyn83w/eva_temp.c:11:Warning:
  out of bounds write. assert \valid(tab + 2);
[kernel] temp_files/tmphpdyn83w/eva_temp.c:11:Warning:
  all target addresses were invalid. This path is assumed to be dead.
\end{lstlisting}

\begin{figure}[htbp]
\begin{lstlisting}[style=cstyle]
/*@ requires x >= INT_MIN/2 && x <= INT_MAX/2 && y >= INT_MIN && y <= INT_MAX;
  @ requires \valid(tab + (0..2));
  @ requires y <= \length(tab) - 5;
  @ assigns tab[2], tab[y+4];
*/
int testme(int x, int y, int tab[]) {
  int v;
  int *pt;
  int **ppt;
  v = x * 2;
  v = v - y;
  pt = &tab[2];
  tab[2] = x;
  ppt = &pt;
  pt = pt + 1;
  *ppt = &tab[2];
  *(pt + v) = y;
  if (tab[y + 4] > 5)
    return 1;
  return 0;
}
\end{lstlisting}
\caption{Alias5 specification generated with an EVA report as context}
\label{code:alias5_eva}
\end{figure}

Notice how the alarms are utilised by the LLM in generating preconditions that ensure that the program avoids both integer overflow and index-out-of-bounds errors. These precise runtime-aware annotations are largely unique to annotations generated with EVA as the reports provide a level of direction and precision absent in the baseline prompt.

\subsubsection{EVA Tunnel-vision}
In examining the impact of incorporating EVA reports into the specification generation process, it becomes apparent that the LLM prioritises the analysis of the EVA report over the direct reasoning about the program's implementation. This observation emerges distinctly when analysing the CoT produced by the LLM. For example, the following CoT excerpt is from the \textbf{LabelsTriTyp} program in \textbf{eva\_set}

\begin{lstlisting}[style=cotstyle]
The provided C program is a function `testme` that takes three integer inputs `Side1`, `Side2`, and `Side3` and uses these inputs to calculate and return a value `triOut`.

From the EVA report, we can see several warnings about potential signed overflow in the program. This occurs when the sum of two sides of the triangle is close to the maximum value an integer can hold (`2147483647`), leading to a possible overfl
\end{lstlisting}
\label{listing:cot2}

In this case we notice that while the specification in \textbf{eva\_set} was able to infer that this program was a triangle classifier, its focus was clearly taken by the provided EVA report.

This observation raises concerns about the LLM's ability to balance the input from static analysis tools like EVA that provide clear direction for the LLM. While the safety and domain boundaries are well-captured, the essence of what the program is designed to do can sometimes be overshadowed by the focus on avoiding runtime errors and handling edge cases as dictated by the EVA report.

\subsection{Implementation vs. Intent}
What should a specification generation tool do if given a buggy program? Symbolic tools, such as QuickSpec \cite{quickspec}, will simply generate some odd specification to which the buggy program adheres. The user might find this surprising, but it might not be obvious that it is due to a bug. Ideally, we would like a specification synthesis tool to generate a specification for what the program \textit{should} do rather than blindly following what is already implemented. LLMs can sometimes help with exactly this: as illustrated in \cite{johansson2023exploring}, GPT-4 produced the intended properties for a functional program even with a subtle bug, by inferring intent from function names rather than details of the implementation. The names provide the LLM with a high-probability cue to the program intent, allowing it to compensate for the bug, see \cite{mccoy2023embers} for a general discussion on this.

In the reminder of this section, we perform a small-scale experiment to investigate the question as to when GPT-4 follows the intent of the program or the implementation. We accomplish this by generating specifications from programs in \textbf{mutated\_set} which contains 8 buggy programs.

\subsubsection{Baseline prompt - intent is prioritised}
While the specifications generated with the baseline prompt were in general lower quality and less likely to infer intent, we did indeed observe that \textbf{if the intent is inferred} either through a function name or from the implementation then the LLM is more likely to create a specification based on that intent even if the buggy implementation contradicts it. One clear example of this is the mutated Tritype function shown in Fig.\ref{code:baseline_mutated_tritype}. In this example, the implementation and specification are disconnected from each other as the LLM's inferred intent took priority over small yet semantically important implementation details.

\begin{figure}[htbp]
\begin{lstlisting}[style=cstyle]
/*@
  ...
  @ behavior not_triangle:
  @   assumes i == 0 || j == 0 || k == 0 || i+j <= k || j+k <= i || i+k <= j;
  @   ensures \result == 4;
  ...
*/
int tritype(int i, int j, int k){
    ...
      @@if ((i+j <= k) || (i+k <= i) || (i+k <= j))@@
	        type_code = 4;
    ...
}
\end{lstlisting}
\caption{Mutated Tritype where specification and implementation clearly do not agree}
\label{code:baseline_mutated_tritype}
\end{figure}

On the other hand, the code for calculating the Levenshtein distance shown in Fig.\ref{code:baseline_mutated_levenshtein} shows us that rather than generating loop annotations from a nuanced understanding of the intent, the LLM instead only considered the implementation of nearby loop implementation.

\begin{figure}[htbp]
\begin{lstlisting}[style=cstyle]
...

int levenshtein(char *s1, char *s2) {
    /*@ loop invariant 0 <= x <= len1;
      @ loop assigns matrix[0][0 .. len1];
    */
    for (int x = 0; x <= len1; x++) @@matrix[0][x]@@ = x;

    /*@ loop invariant 0 <= y <= len2;
      @ loop assigns matrix[0 .. len2][0];
    */
    for (int y = 0; y <= len2; y++) @@matrix[y][0]@@ = y;
}
\end{lstlisting}
\caption{Mutated Levenshtein distance with typos highlighted in red}
\label{code:baseline_mutated_levenshtein}
\end{figure}

When an obvious bug in the code is detected, the LLM responds in one of two primary ways. The more common response is that the LLM disregards the error and generates a specification using inferred intent from other language cues in the program. This lines up with observations made by Gu et al.~\cite{gu2024counterfeit} where they point out that the LLMs they used often had limited understandings of the semantics of the buggy program and could easily be tricked by cues such as variable or function names. 
On less frequent occasions the LLM might actively repair the code during the specification process, contrary to the instructions to not modify the C code. Quite likely the LLM might have seen very similar programs during training, and the corrected version is simply a more likely continuation.

\subsubsection{Prompts with additional annotations}
Adding information from the formal methods tools will not help much in the case of buggy programs. The inclusion of an EVA report continues to steer the LLM's towards specifications focusing on value domains, ignoring any underlying intent of the program, much like observations from bug-free scenarios.
Similarly for Pathfinder, input/output pairs did not consistently aid the LLM in recognising buggy code within a program, even when an oracle was provided specifying which test cases failed. Instead, the specifications largely followed cues from names in the program, as in the baseline case.


\section{Related Work}


There is an ever-growing body of work exploring the opportunities of combining LLMs with various tools for formal-methods and theorem proving. 
In the domain of proof-assistants in particular, a few works have explored the task of synthesising properties or lemmas using neural methods \cite{urban2020,rabe2021,johansson2023exploring} with varying success. More focus has been on creating models for generating proofs, with applications to most mainstream proof assistants like Isabelle/HOL, Coq and Lean \cite{yang2023leandojo,thor,baldur,tactician}. Recent developments of proof co-pilots for Lean are ongoing, aiming to make next-tactic suggestions for the user while creating proof scripts \cite{welleck2023llmstep,song2024towards}.

For contract-based verification tools, the majority of research on annotation generation has been centred on generating loop invariants (e.g., \cite{kamath2023finding,wehrheimAndCoFASE24}) needed to verify correctness of programs wrt.\ a \emph{given} functional specification. Very recently (27/05/2024), there appeared work on arXiv on generating assertions by LLMs with previously encountered errors added to the input \cite{mugnier2024laurel}, to support search for correctness proofs. With the latter, we have in common that some result from symbolic analysis is used when, iteratively, prompting LLMs. However, all of these works assume a functional specification to be given. Our research, on the other hand, addresses largely the problem of \emph{inferring} a functional specification for a program. This ability to generate specifications based on inferred intent could be particularly useful for developer productivity and user adoption. Silva et.al.~\cite{silva2024leveraging} describe how LLMs can be leveraged to help less experienced users craft formal specifications and prove their correctness in the Dafny programming language.

\section{Conclusion}

In this work, we explore how two types of symbolic analyses, symbolic execution based test generation and abstract interpretation based value analysis, enhance the capability of an LLM to extract the intent of C programs and render it as ACSL specifications. The experiments show that without symbolic analysis, the LLM tends to generate lower level annotations reflecting implementation details, like \texttt{assert} and \texttt{assigns} statements. In contrast, the addition of symbolic analyses' output to LLM prompts increases the LLM's capabilities to generate specifications which capture the \emph{intent} of the code, in particular pre- and postconditions. The experiments also show that the specification generation of high-level (intent related) properties was rather robust to bug-introducing mutations of the code. This highlights the value of the overall approach. If generated specifications of buggy code reflect the intended behaviour better than the actual behaviour, then specification generation of this kind can be of great help for revealing bugs.

We note that various limitations of the concrete tool chain we used are not inherent to the approach. For instance, for Pathcrawler sometimes outputs a very small number of input/output pairs, due to its strong focus on path coverage. In such cases, the test cases cannot contribute to more high-level specifications. Also, Pathcrawler-generated tests do not reflect pre- and post states. Both limitations can be mitigated by further development on the Pathcrawler side, or by using a different test generation facility.

Another limitation of this work is the non-determinism of the state-of-the-art LLM services like GPT-4. This makes the results of any experimentation of a tool chain involving such an LLM not fully reproducible.
We share this problem with all work that uses these services in experiments. In the future, we will investigate the usage of LLM frameworks which we can deploy and control locally. Currently, these frameworks seem less competitive for our task, but that may change. Moreover, fine-tuning pre-trained models for our purpose is another direction to further investigate.

This work is aligned to the vision of trustworthy triple copiloting of implementations, tests, and specifications (`TriCo'), as co-outlined by two of the authors in \cite{ahrendt2022trico}. However, in that vision paper, we focused on the bilateral relations of the three artefacts, whereas here, when adding the Pathcrawler output, we use implementations and tests at once when generating specifications. More generally, we see our work as a contribution to the more general aim of combining the complementary strengths of machine learning and exact analyses for effective and reliable development of trustworthy software. 






\bibliographystyle{splncs04}
\bibliography{references}

\newpage
\appendix



\section{Baseline Prompt}
\label{PromptBaseline}
Prompt used for generating ACSL without any additional context.
\begin{lstlisting}[style=cotstyle]
You are a LLM that takes the following inputs and returns a C program annotated with ACSL annotations.

Inputs:
1. A C program with no ACSL annotations

GOALS:
1. Describe any abstract properties that could be represented as ACSL annotations
2. Generate ACSL annotations based on your analysis of the program
3. Returning a program with no annotation is not a valid solution
4. Do not edit the C code, only add annotations
5. Make sure to describe your thought process behind the annotations
6. Do not skip any code in the returned solution to make it shorter.
7. If you break any of these rules then my family will disown me.

ANNOTATION EXAMPLES:

Examples 1 (single annotation):
/*@ requires low >= 0 && high <= 9; */

Example 2 (block annotation style):
//Only use this style for function headers. Do not use blocks for multiple annoations in the function body
/*@
  @ requires low >= 0 && high <= 9;
  @ requires elem >= 0 && elem <= 9;
*/

Example 3 (loops):
/*@
  @ loop invariant low <= high;
  @ loop variant high - low;
*/
while(low <= high)

Example 4 (loop assigns) (loop assigns must be placed before loop variant):
/*@
  @ loop invariant i >= 0 && i <= 3;
  @ loop assigns fa;
  @ loop variant 3 - i;
*/
while(low <= high)

Example 5 (assigns must be in scope):
//This is VALID because x is a parameter that the function contract can see
{valid_assigns}

// this is NOT VALID because x is in the function body and can not be seen by the contract
{invalid_assigns}


FORMAT INSTRUCTIONS:

First describe your reasoning behind the added annotations

Return the annotated c code wrapped in markdown
```c
...
```

START OF INPUT:

```c
{program}
```
\end{lstlisting}
\section{Pathcrawler Augmented Prompt}
\label{PromptPC}
The edited instructions for generating specifications with Pathcrawler input/outputs.
\begin{lstlisting}[style=cotstyle]
...
GOALS:
1. Describe any abstract properties of the input program that could be represented as ACSL annotations
2. Analyze the pathcrawler CSV and describe any patterns that you see that could help you understand the behaviors of the program based on given input/output pairs
3. Describe how these behaviors could be used into creating ACSL annotations
4. Generate ACSL annotations based on your analysis of the program and take special account of the properties described when analyzing the Pathcrawler CSV file
5. Returning a program with no annotation is not a valid solution
6. Do not edit the C code, only add annotations
7. Make sure to describe your thought process behind the annotations
8. Do not skip any code in the returned solution to make it shorter.
9. If you break any of these rules then my family will disown me.
...
START OF INPUT:

Program:
```c
{program}
```

PathCrawler Output:
{csv}
\end{lstlisting}

\section{EVA Augmented Prompt}
\label{PromptEVA}
The edited instructions for generating specifications with EVA reports.
\begin{lstlisting}[style=cotstyle]
...
1. Describe any abstract properties of the input program that could be represented as ACSL annotations
2. Analyze the Eva report and describe how the results could be used in generating ACSL annotations
3. Generate ACSL annotations based on your analysis of the program and take special account of the properties described when analyzing the Eva report
4. Returning a program with no annotation is not a valid solution
5. Do not edit the C code, only annotations
6. Make sure to describe your thought process behind the annotations
7. Do not skip any code in the returned solution to make it shorter.
8. If you break any of these rules then my family will disown me.
...
START OF INPUT:

Program:
```c
{program}
```

Eva Report:
{eva}
...
\end{lstlisting}

\section{Baseline Binary Search}
\label{BaselineBsearch}
Specification for Binary Search generated with the baseline prompt.

\begin{lstlisting}[style=cstyle]
/*@
  @ requires \valid(A + (0..9)) && \valid(&elem);
  @ ensures \result == 0 || \result == 1;
  @ assigns \nothing;
*/
int testme( int A[10], int elem) {
  int low, high, mid, ret ;
  low = 0 ;
  high = 9 ;
  ret = 0 ;
  /*@
    @ loop invariant high >= low;
    @ loop assigns low, high, mid, ret;
    @ loop variant high - low;
  */
  while( ( high > low ) )
    { mid = (low + high) / 2 ;

      if( elem == A[mid] )
         ret = 1;
      if( elem > A[mid] )
        low = mid + 1 ;
      else
        high = mid - 1;
    }
  mid = (low + high) / 2 ;

  if( ( ret != 1)  && ( elem == A[mid]) )
    ret = 1;

  return ret ;
}
\end{lstlisting}

\section{Preliminary Baseline Binary Search}
\label{PrelimBaselineBsearch}
Specification generated with an earlier verison of GPT-4

\begin{lstlisting}[style=cstyle]
/*@
  @ requires \valid(A + (0..9));
  @ requires \forall integer i, j; 0 <= i < j < 10 ==> A[i] <= A[j]; // Array is sorted
  @ assigns \nothing;
  @ ensures \result == 1 <==> \exists integer i; 0 <= i < 10 && A[i] == elem;
*/
int testme( int A[10], int elem) {
  int low, high, mid, ret ;

  /*@
    @ assigns low;
  */
  low = 0 ;

  /*@
    @ assigns high;
  */
  //@assert true;
  high = 9 ;

  /*@
    @ assigns ret;
  */
  ret = 0 ;

  /*@
    @ loop invariant 0 <= low <= high <= 9;
    @ loop assigns low, high, mid, ret;
    @ loop variant high - low;
  */
  while( ( high > low ) )
    {
      /*@
        @ assigns mid;
      */
      mid = (low + high) / 2 ;

      /*@
        @ assigns ret;
      */
      if( elem == A[mid] )
         ret = 1;

      /*@
        @ assigns low;
      */
      if( elem > A[mid] )
        low = mid + 1 ;
      else
      /*@
        @ assigns high;
      */
        high = mid - 1;
    }

  /*@
    @ assigns mid;
  */
  mid = (low + high) / 2 ;

  /*@
    @ assigns ret;
  */
  if( ( ret != 1)  && ( elem == A[mid]) )
    ret = 1;

  /*@
    @ assigns \nothing;
  */
  return ret ;
}
\end{lstlisting}

\section{LabelsTritype Augmented with an Eva Report}
\label{LabelsTritypeEVA}
LabelsTritype is a version of Tritype that is meant to test the labeling feature of Pathcrawler. The following is a specification generated for LabelsTritype with an Eva report added.

\begin{lstlisting}[style=cstyle]
/*@ requires Side1 > 0 && Side2 > 0 && Side3 > 0;
  @ ensures \result >= 1 && \result <= 4;
  @ assigns \nothing;*/
int testme (int Side1, int Side2, int Side3){
   int triOut;
   /*@ assert Side1 <= 2147483647 && Side2 <= 2147483647 && Side3 <= 2147483647; */
   if (Side1 <= 0 || Side2 <= 0 || Side3 <= 0)
   {
      triOut = 4;
      return (triOut);
   }
   triOut = 0;
   if (Side1 == Side2){ triOut = triOut + 1;}
   if (Side1 == Side3){ triOut = triOut + 2;}
   if (Side2 == Side3){triOut = triOut + 3;}
   if (triOut == 0)
   {
     /*@ assert Side1+Side2 <= 2147483647 && Side1+Side3 <= 2147483647 && Side2+Side3 <= 2147483647; */

     if (Side1+Side2 <= Side3 || Side2+Side3 <= Side1 || Side1+Side3 <= Side2){triOut = 4;}
     else{triOut = 1;}
     return (triOut);
   }
   if (triOut > 3){triOut = 3;}
   else{
     /*@ assert Side1+Side2 <= 2147483647 && Side1+Side3 <= 2147483647 && Side2+Side3 <= 2147483647; */
     if (triOut == 1 && Side1+Side2 > Side3){triOut = 2;}
     else{
       if (triOut == 2 && Side1+Side3 > Side2){triOut = 2;}
       else{
         pathcrawler_label(triOut == 3);
         pathcrawler_label(triOut != 3);
         pathcrawler_label(Side2+Side3 > Side1);
         pathcrawler_label(Side2+Side3 <= Side1);
         if (triOut == 3 && Side2+Side3 > Side1){triOut = 2;}
         else{triOut = 4;}
       }
     }
   }
   return (triOut);
}
\end{lstlisting}

The EVA report provided to the prompt:
\begin{lstlisting}[style=cotstyle]
[kernel] Parsing temp_files/tmpifalg4bf/eva_temp.c (with preprocessing)
[kernel:typing:implicit-function-declaration] temp_files/tmpifalg4bf/eva_temp.c:20: Warning:
  Calling undeclared function pathcrawler_label. Old style K&R code?
[eva] Option -eva-precision 7 detected, automatic configuration of the analysis:
  option -eva-min-loop-unroll set to 0 (default value).
  option -eva-auto-loop-unroll set to 256.
  option -eva-widening-delay set to 4.
  option -eva-partition-history set to 1.
  option -eva-slevel set to 250.
  option -eva-ilevel set to 128.
  option -eva-plevel set to 300.
  option -eva-subdivide-non-linear set to 140.
  option -eva-remove-redundant-alarms set to true (default value).
  option -eva-domains set to 'cvalue,equality,gauges,octagon,symbolic-locations'.
  option -eva-split-return set to 'auto'.
  option -eva-equality-through-calls set to 'formals' (default value).
  option -eva-octagon-through-calls set to true.
[eva] Splitting return states on:
[eva] Analyzing an incomplete application starting at testme
[eva] Computing initial state
[eva] Initial state computed
[eva:initial-state] Values of globals at initialization

[kernel:annot:missing-spec] temp_files/tmpifalg4bf/eva_temp.c:20: Warning:
  Neither code nor specification for function pathcrawler_label, generating default assigns from the prototype
[eva] using specification for function pathcrawler_label
[eva:alarm] temp_files/tmpifalg4bf/eva_temp.c:67: Warning:
  signed overflow. assert Side1 + Side2 <= 2147483647;
[eva:alarm] temp_files/tmpifalg4bf/eva_temp.c:69: Warning:
  signed overflow. assert Side1 + Side3 <= 2147483647;
[eva:alarm] temp_files/tmpifalg4bf/eva_temp.c:71: Warning:
  signed overflow. assert Side2 + Side3 <= 2147483647;
[eva:alarm] temp_files/tmpifalg4bf/eva_temp.c:100: Warning:
  signed overflow. assert Side1 + Side2 <= 2147483647;
[eva:alarm] temp_files/tmpifalg4bf/eva_temp.c:112: Warning:
  signed overflow. assert Side1 + Side3 <= 2147483647;
[eva:alarm] temp_files/tmpifalg4bf/eva_temp.c:124: Warning:
  signed overflow. assert Side2 + Side3 <= 2147483647;
[eva] done for function testme
[eva] ====== VALUES COMPUTED ======
[eva:final-states] Values at end of function testme:
  triOut in {1; 2; 3; 4}
  __retres in {1; 2; 3; 4}
[eva:summary] ====== ANALYSIS SUMMARY ======
  ------------------------------
  1 function analyzed (out of 1): 100% coverage.
  In this function, 73 statements reached (out of 73): 100% coverage.
  ------------------------------
  Some errors and warnings have been raised during the analysis:
    by the Eva analyzer:      0 errors    0 warnings
    by the Frama-C kernel:    0 errors    2 warnings
  ------------------------------
  6 alarms generated by the analysis:
       6 integer overflows
  -------------------------------
  No logical properties have been reached by the analysis.
\end{lstlisting}

\section{Mutated Levenshtein}
\label{code:code_mutated_levenshtein}
Mutated Levenshtein distance algorithm with typos highlighted in red
\begin{lstlisting}[style=cstyle]
#include <string.h>

/*@ requires \true;
  @ ensures \result <= x && \result <= y && \result <= z;
  @ assigns \nothing;
*/
int min(int x, int y, int z) {
    if (x < y) return (x < z) ? x : z;
    else return (y < z) ? y : z;
}
/*@ requires \valid(s1) && \valid(s2);
  @ ensures \result >= 0;
  @ assigns \nothing;
*/
int levenshtein(char *s1, char *s2) {
    int len1 = strlen(s1), len2 = strlen(s2);
    int matrix[len1 + 1][len2 + 1];

    /*@ loop invariant 0 <= x <= len1;
      @ loop assigns matrix[0][0 .. len1];
    */
    for (int x = 0; x <= len1; x++) @@matrix[0][x]@@ = x;

    /*@ loop invariant 0 <= y <= len2;
      @ loop assigns matrix[0 .. len2][0];
    */
    for (int y = 0; y <= len2; y++) @@matrix[y][0]@@ = y;
    /*@ loop invariant 1 <= x <= len1;
      @ loop assigns matrix[1 .. len1][1 .. len2];
    */
    for (int x = 1; x <= len1; x++) {
        /*@ loop invariant 1 <= y <= len2;
          @ loop assigns matrix[x][1 .. len2];
        */
        for (int y = 1; y <= len2; y++) {
            int cost = (s1[x - 1] == s2[y - 1]) ? 0 : 1;
            matrix[x][y] = min(matrix[x - 1][y] + 1, matrix[x][y - 1] + 1, matrix[x - 1][y - 1] + cost);
        }
    }
    return matrix[len1][len2];
}
\end{lstlisting}

\section{BugKPath Baseline}
\label{code:bugkpath_baseline}
\begin{lstlisting}[style=cstyle]
/*@ requires \valid(msg + (0 .. len-1)) && \valid(buffer + (0 .. BUFSZ-1));
  @ requires len >= 0;
  @ assigns buffer[0 .. BUFSZ-1];
  @*/
void testme (char *msg, int len, char *buffer)
{
  int i;
  int j;
  int limit = BUFSZ - 1;
  /*@ loop invariant 0 <= i <= len;
    @ loop variant len - i;
    @*/
  for (i = 0; i < len; ) {
    /*@ loop invariant 0 <= j <= limit;
      @ loop assigns buffer[0 .. limit];
      @ loop variant limit - j;
      @*/
    for (j = 0; i < len && j < limit; ){
    ...
    }
  }
}
\end{lstlisting}

\section{Apache Pathcrawler}
\label{ApachePC}
Specification generated for the Apache program with Pathcrawler input/output pairs as additional context

\begin{lstlisting}[style=cstyle]
#include "apache.h"

/*@
  @ requires \valid(s);
  @ ensures \result >= 0;
*/
unsigned mystrlen(char *s)
{
  int i;
  i = 0;
  /*@
    @ loop invariant \valid(s+i);
    @ loop assigns i;
  */
  while (s[i] != EOS)
    ++i;
  return i;
}

/*@
  @ requires \valid(s1) && \valid(s2) && n >= 0;
  @ ensures \result <= 0 || \result == 0 || \result > 0;
*/
int mystrncmp (const char *s1, const char *s2, int n)
{
  int i;
  int retval;
  i = 0;
  /*@
    @ loop invariant \valid(s1+i) && \valid(s2+i) && i <= n;
    @ loop assigns i, retval;
  */
  do {
    retval = s1[i] - s2[i];
    if (i >= n-1) return retval;
    if (retval != 0) return retval;
    if (s1[i] == EOS) return 0;
    i++;
  } while (1);
}
/*@
  @ requires scheme >= 0 && \valid(uri);
  @ assigns uri[0..TOKEN_SZ];
*/
void testme (char *uri, int scheme)
{
  int cp;
  int c,i;
  char LDAP[5]={"ldap"};
  char *token[TOKEN_SZ];
  if (scheme == 0
      || mystrlen(uri) < scheme) {
    return;
  }
  cp = scheme;

  if (uri[cp-1] == '/') {

    while (uri[cp] != EOS
           && uri[cp] != '/') {
      ++cp;
    }
    if (uri[cp] == EOS || uri[cp+1] == EOS) return;
    ++cp;
    scheme = cp;
    if (mystrncmp(uri, LDAP, LDAP_SZ) == 0) {
      c = 0;
      token[0] = uri;

      while (uri[cp] != EOS
             && c < TOKEN_SZ) {
        if (uri[cp] == '?') {
          ++c;
          token[c] = uri + cp + 1;
          uri[cp] = EOS;
        }
        ++cp;
      }
      return;
    }
  }
  return;
}
\end{lstlisting}

The Pathcrawler Input/Output pairs provided. Notice how no output is provided. Since Apache is a void function with side-effects, the input output pairs say nothing about the program.

\begin{lstlisting}
input_scheme,input_uri[0],...,input_uri[14],output,verdict
1,47,47,0,0,0,0,0,0,0,0,0,0,0,0,0,,unknown
1,47,58,0,0,0,0,0,0,0,0,0,0,0,0,0,,unknown
2,108,47,47,47,0,0,0,0,0,0,0,0,0,0,0,,unknown
1,47,0,0,0,0,0,0,0,0,0,0,0,0,0,0,,unknown
5,108,100,97,112,47,47,63,0,0,0,0,0,0,0,0,0,unknown
5,108,100,97,112,47,47,63,47,63,63,0,0,0,0,0,,no_extra_coverage
5,108,100,97,112,47,47,63,63,63,0,0,0,0,0,0,,no_extra_coverage
0,0,0,0,0,0,0,0,0,0,0,0,0,0,0,0,,unknown
422214939,0,0,0,0,0,0,0,0,0,0,0,0,0,0,0,,unknown
1,47,47,47,0,0,0,0,0,0,0,0,0,0,0,0,,unknown
1,58,0,0,0,0,0,0,0,0,0,0,0,0,0,0,,unknown
5,108,100,97,112,47,47,63,63,63,47,0,0,0,0,0,,unknown
5,108,100,97,112,47,47,63,63,47,63,0,0,0,0,0,,no_extra_coverage
4,108,100,97,47,47,47,0,0,0,0,0,0,0,0,0,,unknown
5,108,100,97,112,47,47,47,0,0,0,0,0,0,0,0,,unknown
422214939,47,0,0,0,0,0,0,0,0,0,0,0,0,0,0,,unknown
5,108,100,97,112,47,47,47,63,63,63,0,0,0,0,0,,no_extra_coverage%
\end{lstlisting}

\end{document}